\documentclass[12pt,nofootinbib,showkeys,prd,aps]{revtex4-1}

\usepackage[brazil, english]{babel}
\usepackage[utf8]{inputenc}
\usepackage{graphicx}
\usepackage{epsfig}
\usepackage{amssymb}
\usepackage{amsmath} 
\usepackage{slashed}
\usepackage{physics}
\usepackage{subcaption}
\usepackage{caption}
\usepackage[unicode=true,bookmarks=false,breaklinks=false,pdfborder={0 0 1},colorlinks=true]
 {hyperref}
\hypersetup{
 citecolor=blue,linkcolor=blue,urlcolor=blue}

\graphicspath{%
    {converted_graphics/}% inserted by PCTeX
    {/}% inserted by PCTeX
}
\begin{document}

%\date{}
\title{Mechanical properties of the proton from a deformed AdS holographic model}
\author{Ayrton Nascimento$^{1}$}
\email{ayrton@pos.if.ufrj.br}
\author{Henrique Boschi-Filho$^1$}
\email{boschi@if.ufrj.br}  
\affiliation{$^1$Instituto de F\'\i sica, Universidade Federal do Rio de Janeiro, 21.941-909, Rio de Janeiro, RJ, Brazil}

\begin{abstract}
We study the gravitational form factors of the proton and some of its mechanical properties. We use a holographic model based on the AdS/CFT correspondence, in which a deformation in the anti-de Sitter background geometry is considered. By describing the proton as a Dirac field in this background, we numerically evaluate the gluon contribution of its gravitational form factors $A$ and $C$ from its energy-momentum tensor. A comparison of our numerical results  with respect to some lattice QCD results and previous results in holography is made. In general, a good agreement is found. We also evaluate the term $D$ and make use of it to compute the pressure and shear distributions in the system, which result in a stable composed particle interpretation consistent with the von Laue stability condition. The energy distribution in the system is also obtained. Internal forces are investigated to support this picture. We are also able to compute the radii associated with these distributions in the proton.
\end{abstract}


\maketitle

%\tableofcontents

%\begin{multicols}{2}

\section{Introduction}
\label{Intro}

\hspace{5mm} The proton, one of the building blocks of nuclear matter and, consequently, ordinary stuff, has been known for more than one century. However, some of its properties remain at least partially unknown. This is due mainly to its internal structure with non-trivial contributions from its constituents, the quarks and gluons. One way to address this endeavor is to look at the Energy Momentum Tensor's (EMT) gravitational form factors (GFFs), which encode information regarding the distribution of its mass or momentum, spin, shear, and pressure. This question naturally expands to nucleons and other hadrons \cite{Burkert:2018bqq,
Polyakov:2018zvc,Deur:2018roz,Lorce:2018egm,Burkert:2023wzr}. In principle, GFFs can be probed by the gravitational interaction of quarks and gluons with the graviton, but this constitutes a challenging experimental effort because of the weakness of the gravitational interaction at the quantum level. This was proposed long ago by Gell-Mann \cite{Gell-Mann:1962yej}. Recently, experimental extractions of GFFs have been performed indirectly, for instance, accessing them through deep virtual Compton scattering data \cite{Burkert:2018bqq,Burkert:2021ith}. The future Electron Ion Collider \cite{Accardi:2012qut} looks to further extend such experimental investigations using scattering processes to extract GFFs and to investigate the nucleon structure and its properties. On the theory side, recent developments in lattice QCD, such as \cite{Alexandrou:2018xnp,Shanahan:2018pib,Shanahan:2018nnv,Hackett:2023rif}, explored proton GFFs and some of its mechanical properties, such as pressure and shear distributions (see also \cite{Dehghan:2025ncw, Goharipour:2025lep, Goharipour:2025yxm}). 

As non-perturbative aspects of QCD related to its strong coupling regime, such properties of the nucleon are also well suited for holographic models based on the gauge/gravity duality \cite{Maldacena:1997re,Gubser:1998bc,Witten:1998qj}, which relates a classical weakly coupled gravity theory on a curved five dimensional anti-de Sitter space to a dual strongly coupled gauge theory on flat Minkowski space. In this context, the works of \cite{Abidin:2008ku,Abidin:2008hn,Brodsky:2008pf, Abidin:2008sb,
Abidin:2009hr, Anderson:2014jia, Chakrabarti:2015lba, Mondal:2015fok, Mondal:2017lph, Kumar:2017dbf, Watanabe:2018owy, 
Mamo:2019mka,Aliev:2020aih, Chakrabarti:2020kdc,  Mamo:2021krl, Karapetyan:2021crv, Liu:2022out,Allahverdiyeva:2023fhn, Mamedov:2024tth,
Deng:2025fpq, Sain:2025kup} paved the way for studies of nucleon and hadronic GFFs and some of their mechanical properties such as those mentioned above using holographic models for QCD, also known as AdS/QCD models. Among these models, we mention the hard wall  \cite{Polchinski:2001tt, Boschi-Filho:2002xih,Boschi-Filho:2002wdj, deTeramond:2005su, Erlich:2005qh, Boschi-Filho:2005xct}, the soft wall \cite{Karch:2006pv, Brodsky:2006uqa, 
Erdmenger:2007cm,
Brodsky:2014yha} (see also \cite{Abidin:2008hn,Abidin:2008sb,Abidin:2008ku,Abidin:2009hr,Mamo:2019mka,Mamo:2021krl,Braga:2011wa}), which are classified as bottom-up models, as well as some top-down models such as the $D4-D8$ Sakai-Sugimoto \cite{Sakai:2004cn,
Sakai:2005yt, 
Hata:2007mb, Fujita:2022jus,
    Fujii:2024rqd, Sugimoto:2025btn}. 
The role of confinement in the soft wall model inspired \cite{Andreev:2006ct} another bottom-up AdS/QCD model known as the deformed AdS  \cite{FolcoCapossoli:2019imm,
FolcoCapossoli:2020pks,
MartinContreras:2021yfz,
Contreras:2021epz}. {A comparison remarking similarities and differences of the soft-wall and deformed AdS models appeared recently in \cite{Nascimento:2023dzx}. The soft-wall model considers the pure AdS space and the introduction of the exponential of a dilaton field in the action \cite{Karch:2006pv}. In the case of fermions it is also necessary to introduce a potential in the mass term to obtain discrete mass spectra \cite{Abidin:2008hn,Abidin:2008ku}. On the other hand, the deformed AdS model considers an exponential with the dilaton field in the metric, which is then only asymptotically AdS and there is no extra dilaton in the action \cite{FolcoCapossoli:2019imm,FolcoCapossoli:2020pks}. In fact, in the soft-wall model, the dilaton-like term present in the fields' action appears through an exponential that factors out of the equations of motion, making no effect on the fermion field dynamics. It also does not affect the fields' mass spectra. Introducing a $z$ dependent mass term in the action modifies the fields' dynamics and, together with the dilaton, leads to a discrete spectrum of eigenvalues. Regarding the deformed AdS model, in which the dilaton-like term goes into the line element of the asymptotic AdS background, there is no need for an additional mass term proportional to the $z$ coordinate in the Dirac action, leaving it more natural in the bulk geometry. As a consequence, a discrete spectrum naturally emerges. In addition, Schrödinger-like equations of motion coming from a bulk Dirac action present numerical solutions, as opposed to the analytical solutions present in the soft-wall model in \cite{Abidin:2008hn,Mamo:2019mka}. In fact, as we have shown in \cite{Nascimento:2023dzx} in the case of fermions, in the deformed AdS model, in a quadratic approximation, the equations of motion can be cast in a way that look like those of the soft-wall model, leading to similar analytical solutions, but even in this case their spectra differ by the ground state, which shows the models, in this approximation, may look similar to one another, but are fundamentally different.}

In this work, we explore the proton gravitational form factors by computing the eigenvalues of the nucleon EMT in a deformed AdS background holographic model \cite{ FolcoCapossoli:2019imm,
FolcoCapossoli:2020pks,
MartinContreras:2021yfz,
Contreras:2021epz}. One of them, the $D$ term, allows one to explore mechanical properties of the system such as pressure and shear distributions, as well as internal forces.

This paper is organized as follows. In Section \ref{HM}, we discuss the deformed AdS holographic model used in this work and describe the nucleon as a Dirac field in the background spacetime of this model. Section \ref{GFFs} is devoted to the computation of the gravitational form factors of the nucleon, and the comparison of our results with some lattice QCD data, as well as with other holographic results. From these, in Section \ref{PressureShear}, we explore the pressure, shear, and energy distributions in the nucleon using a dipole approximation to our numerical results, which also allows us to compute the mechanical radius associated with these distributions. Using these results, in Section \ref{Forces}, we investigate the internal forces in the proton, such as normal and tangential forces inside of it. We draw our conclusions and future perspectives in Section \ref{conclusions}. We also include an Appendix \ref{SoftWall} where we briefly review the usual soft wall model for fermions, pinpointing the differences with the deformed AdS model discussed in Section \ref{HM}.

%%%%%%%%%%%%%%%%%%%%%%%%%
%%%%%%  Section  %%%%%%%%
%%%%%%%%%%%%%%%%%%%%%%%%%

\section{Deformed AdS model}\label{HM}

\hspace{5mm} In the deformed AdS model, we consider a five-dimensional anti-de Sitter background deformed by a warp factor $\mathcal{A}(z)$ as follows \cite{Contreras:2021epz, Nascimento:2023dzx} 
\begin{equation}
    ds^{2}=g_{mn}\,dx^m dx^n=e^{2\,\mathcal{A}(z)}(dz^2+\eta_{\mu\nu}\,dy^\mu dy^\nu),\label{eq:ZTmetric}
\end{equation}
where $\{m,n\}$ refers to coordinates in five dimensions, and $\eta_{\mu\nu}$ is the Minkowski metric with signature $(-,+,+,+)$. The warp factor $\mathcal{A}(z)$ is separated into two parts,
{
\begin{equation}\label{WF}
    \mathcal{A}(z)=-\log \left(\frac{z}{R}\right)+\frac{1}{2}k\,z^2,
\end{equation}
where $-\log \left(\frac{z}{R}\right)$, with $R$ the radius of curvature of the asymptotic $\text{AdS}_5$ background, recovered within the limit $z\to0$. The dilaton-like factor of $\frac{1}{2}k\,z^2$, with $k$ a constant having a mass squared dimension, deforms the AdS background} breaking the conformal invariance of the boundary quantum field theory. It was originally chosen to reproduce confining quark-antiquark potentials \cite{Andreev:2006ct}.

%%%%%%%%%%%%%%%%%%%%%%%%
%%%%% Subsection %%%%%%%
%%%%%%%%%%%%%%%%%%%%%%%%

%\subsection{Proton in the deformed AdS background}
\label{Nucleon}

\hspace{5mm} Fermions, such as the proton, are described by a probe Dirac field $\Psi$ in this model in five dimensions \cite{Contreras:2021epz, Nascimento:2023dzx}
\begin{equation}
     S_\psi=\int d z \, d^{4}x \,\sqrt{-g}\,{\bar\Psi}({\slashed{D}}- m_{5})\Psi,\label{Action}
\end{equation}
where $g$ is the determinant of the metric of the bulk geometry \eqref{eq:ZTmetric}, and $m_5$ is the bulk fermionic mass. The operator $\slashed{D}$ is defined by
\begin{equation}
    {\slashed{D}}\equiv  g^{mn}\, e^{a}_{n}\,\Gamma_{a}\left(\partial_{m}+\frac{1}{2}\omega_{m}^{bc}\,\Sigma_{bc}\right),
\end{equation}
where $\Sigma_{bc}=\frac{1}{4}[\Gamma_{b},\Gamma_{c}]$, with Dirac matrices given by $\Gamma^{a}=(\gamma^{\mu},-i\gamma^{5})$. The vielbein $ e^{a}_{n}$ for the deformed background is
\begin{equation}
     e^{a}_{m}=e^{\mathcal{A}( z)}\,\delta^{a}_{m}. 
\end{equation}
The spin connection $ \omega_{m}^{bc}$ is
\begin{equation}
    \omega^{a\,b}_{m}= e^{a}_{n}\,\partial_{m} e^{n\,b}+ e^{a}_{n}\, e^{p\,b}\,\Gamma^{n}_{p\,m}, 
\end{equation}
where $m=0, 1, 2, 3, 5$, and $\Gamma^{n}_{p\,m}$ are the Christoffel symbols
\begin{equation}
    \Gamma^{p}_{m\,n}=\frac{1}{2}\, g^{p\,q}(\partial_{n}\, g_{m\,q}+\partial_{m}\, g_{n\,q}-\partial_{q}\, g_{m\,n}).
\end{equation}
From (\ref{Action}), one can get the following equation of motion
\begin{equation}
    (e^{-\mathcal{A}(z)}\,\gamma^{5}\partial_{z}+e^{-\mathcal{A}(z)}\,\gamma^{\mu}\partial_{\mu}+2\,\mathcal{A}^{\prime}(z)\gamma^{5}- m_{5})\,\Psi=0,\label{eq:Dirac}
\end{equation}
with $\mathcal{A}'(z)$ representing the derivative of $\mathcal{A}(z)$ with respect to the holographic coordinate $z$. {Note that, according to its definition, Eq. \eqref{WF}, $\mathcal{A}(z)$ is dimensionless, and so $\mathcal{A}'(z)$ has dimension of mass, as well as $m_5$, introduced in the fermionic action, Eq. \eqref{Action} (see also the discussion below Eq. \eqref{m5} where we set the value of $m_5$ in GeV, together with other parameters in the model to fit the proton mass.}  

In order to solve Eq. \eqref{eq:Dirac}, we can decompose $\Psi$ into its \textit{left} and \textit{right} chiral components
\begin{equation}
    \Psi(x^{\mu}, z)=\Psi_{_L}(x^{\mu}, z)+\Psi_{_R}(x^{\mu}, z),\label{Pcompts}
\end{equation}
with $\Psi_{L/R}=\frac{1\mp \gamma^5 }{2}\Psi$. Considering that the Kaluza-Klein modes are dual to the chiral components, one can write them as
\begin{equation}
    \Psi_{L/R}(x,z)=\sum_nf^n_{L/R}(x)\chi^n_{L/R}(z).\label{KK}
\end{equation}
Using (\ref{Pcompts}) and (\ref{KK}) in (\ref{eq:Dirac}), we get the following system of coupled equations
\begin{align}
    (\partial_{z}+2\,A^{\prime}(z)\,e^{\mathcal{A}(z)}+m_5\,e^{\mathcal{A}(z)})\,\chi^{n}_L(z)&=M_{n}\,\chi^{n}_R(z),\label{eq:coupledplus}\\
    (\partial_{z}+2\,\mathcal{A}^{\prime}(z)\,e^{\mathcal{A}(z)}-m_5\,e^{\mathcal{A}(z)})\,\chi^{n}_R(z)&=-M_{n}\,\chi^{n}_L(z),\label{eq:coupledminus}
\end{align}
with $M_{n}$ the mass of the $n$th excited {hadronic state in four dimensions. The one corresponding to the ground state $M_{0}$ is to be identified with the proton's mass.} By decoupling these equations and using the following transformations,
\begin{equation}
    \chi^n_{L/R}(z)=\psi^{n}_{L/R}(z)\,e^{-2\mathcal{A}(z)},
\end{equation}
we obtain a Schrödinger-like equation for the fermions
\begin{equation}
    [-\partial^{2}_{z}+V_{L/R}(z)]\, \psi_{L/R}^{n}(z)= M_{n}^{2}\, \psi^{n}_{L/R}(z),\label{SchEq}
\end{equation}
where the corresponding chiral potentials are given by
\begin{equation}
    V_{L/R}(z)= m_{5}^{2}\, e^{2\,\mathcal{A}(z)}\mp \, e^{\mathcal{A}(z)}\, m_{5}\,\mathcal{A}^{\prime}(z).\label{Potential}
\end{equation}

\begin{figure}[ht!]
    \centering
\includegraphics[width=0.8\linewidth]{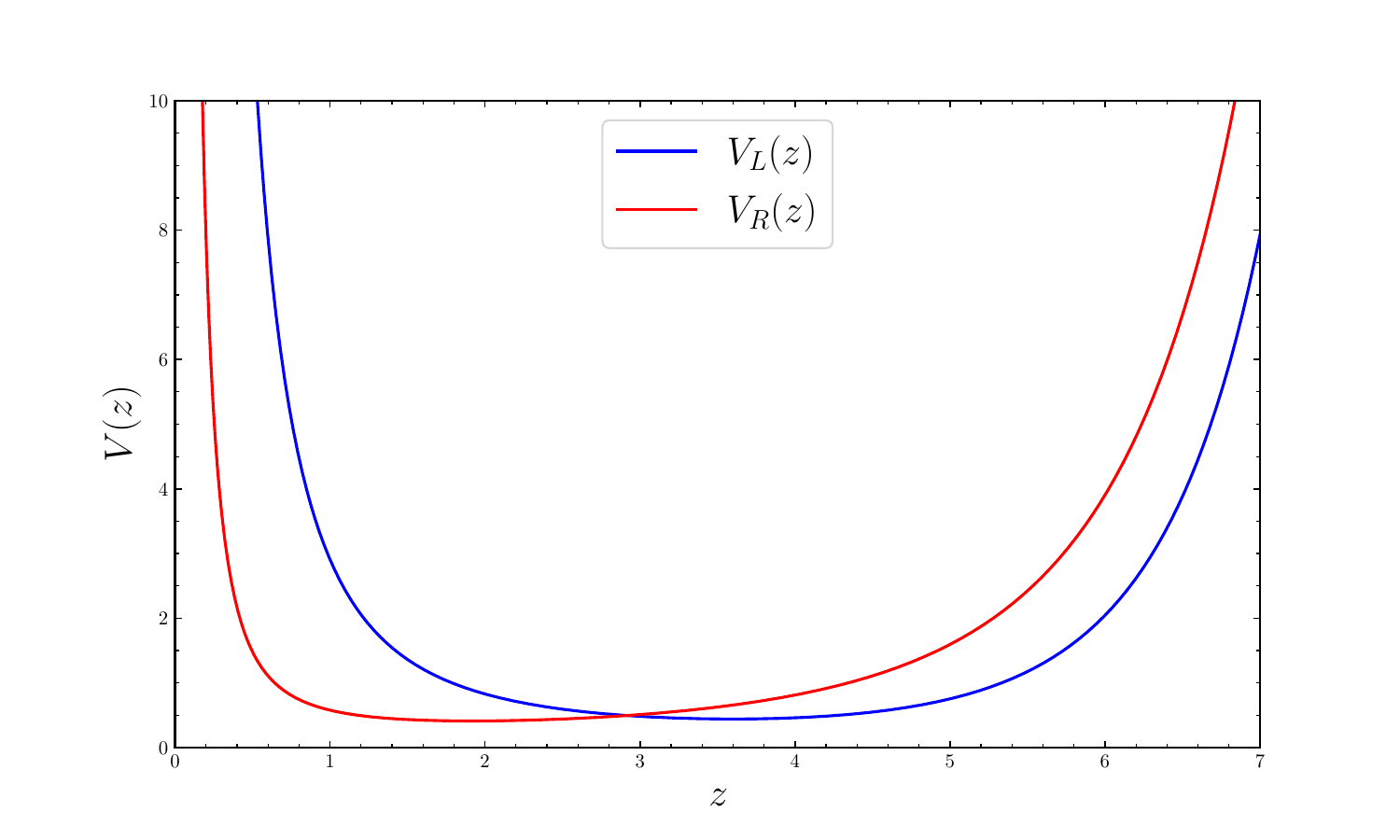}
    \caption{\sl {Confining chiral potentials $V_{L/R}(z)$ which come from \eqref{Potential} with the free parameters given by the particular choice $k=(0.343)^2$$\,\text{GeV}^2$ and $\gamma=1.745$.}}
    \label{Fig:Potentials}.
\end{figure}

In order to model gravitational form factors, we need to solve (\ref{SchEq}) for its ground state. With $\mathcal{A}(z)$ given by (\ref{WF}), the resulting fermionic solution is obtained by numerically solving the Schrödinger-like equation. The goal is to make the ground state mass $M_0$ match the proton mass of $M_p=0.938\,{\text{GeV}}$. We seek for free parameters to achieve this goal. The constant $k$ present in the background warp factor $\mathcal{A}(z)$ could be chosen so that the match to $M_p$ is achieved. {But this is not accomplished. Another free parameter is needed for this match. This is generally resolved by looking at $m_5$ entering \eqref{SchEq} through \eqref{Potential}. In particular, according to the holographic dictionary, in the pure AdS spacetime, $m_5^{\text{AdS}}$ is related to the canonical scaling conformal dimension $\Delta_\text{can}$ of a boundary operator $\mathcal{O}$ through}
\begin{equation}
    R\,m_5^{\text{AdS}}=\Delta_\text{can}-2,
\end{equation}
where $\Delta_\text{can}$ is the fermionic canonical conformal dimension $\Delta=\frac{3}{2}$. With the resulting value of $m_5$, we are unable to reproduce the mass of the proton for any chosen value of $k$ when using it in solving Eq. \eqref{SchEq}. { In fact, in \cite{Georgi:1974wnj,Gross:1973zrg,Gross:1976xt,Jaroszewicz:1982gr} it was shown that the canonical dimension of an operator $\mathcal{O}$ should be modified by introducing an anomalous dimension $\gamma$, implying an effective scaling dimension $\Delta_\text{eff}=\Delta_\text{can}+\gamma-2$. Inspired by this, we consider an anomalous dimension $\gamma$ in $m_5$ such that}
\begin{equation}\label{m5}
    R\,m_5=\Delta_\text{can}+\gamma-2.
\end{equation}
In this way, we can fine tune both parameters, $k$ and $\gamma$, to not only match $M_0$ with the proton mass, but to adjust the predictions of this model for the gravitational form factors of the proton with the corresponding results in the lattice QCD literature, and compare them with previous holographic results. {To solve \eqref{SchEq}, we choose $R=1\,\text{GeV}^{-1}$ and set $z_0=1\times10^{-5}\,\text{GeV}^{-1}$ and $z_\infty=7\,\text{GeV}^{-1}$ as the domain to which constrain the solutions of \eqref{SchEq}. The values of the free parameters that best achieve the desired match between $M_0$ and $M_p$ are found to be $k=(0.343)^2$$\,\text{GeV}^2$, $\gamma=1.745$ and $m_5=1.245\,\text{GeV}$, as illustrated by the masses $M_0$ of the $\psi_L$ and $\psi_R$ solutions in Table \ref{table:2}.
\begin{table}
    \centering
    \begin{tabular}{|c|c|c|}
    \hline
         & $\psi_L$ & $\psi_R$ \\
         \hline
       $M_0$  & $0.937802\,{\text{GeV}}$ & $0.937133\,{\text{GeV}}$\\
       \hline
    \end{tabular}
    \vspace{0.5cm}
    \caption{\sl Masses of the chiral solutions $\psi_L$ and $\psi_R$ for $k=(0.343)^2\,\text{GeV}^2$, $\gamma=1.745$ and $m_5=1.245\,\text{GeV}$.}
    \label{table:2}
\end{table}
The corresponding chiral potentials for \eqref{Potential} present a confining behavior, as shown in Fig. \ref{Fig:Potentials}}.

\begin{figure}[ht!]
    \centering
\includegraphics[width=0.8\linewidth]{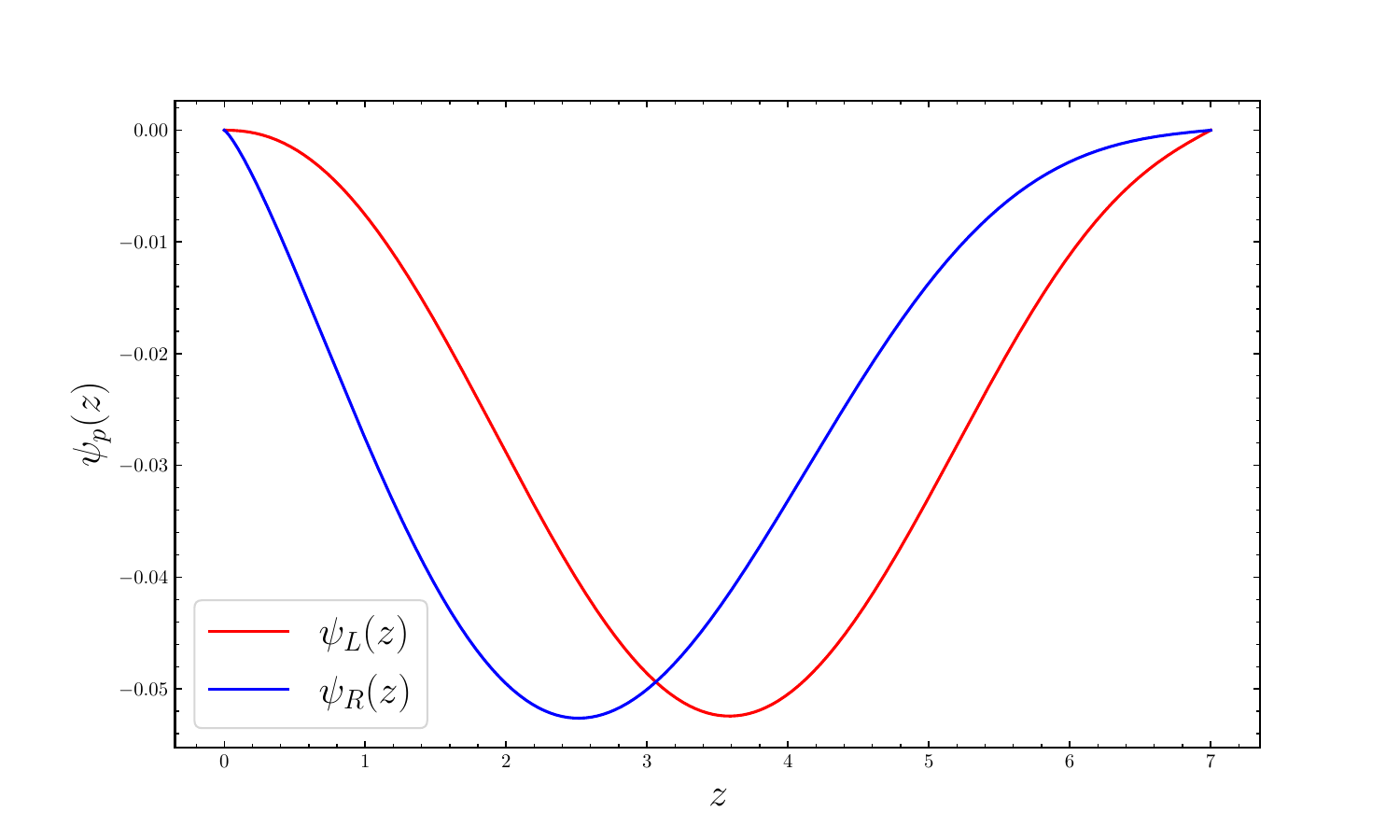}
    \caption{\sl {Ground state solutions $\psi_L(z)$ and $\psi_R(z)$ corresponding to the chiral potentials $V_{L/R}(z)$. We associate these solutions with the proton wave function $\psi_p(z)$ in its ground state in the present holographic model.}}
    \label{Fig:WaveFunctions}
\end{figure}

{The resulting ground state solutions $\psi_L$ and $\psi_R$ corresponding to the eigenvalues in Table \ref{table:2} are depicted in Fig. \ref{Fig:WaveFunctions}. In what follows, we keep all the parameters discussed above fixed.}

%%%%%%%%%%%%%%%%%%%%%%%%%
%%%%%%%  Section %%%%%%%%
%%%%%%%%%%%%%%%%%%%%%%%%%

\section{Gravitational Form Factors}
\label{GFFs}

\hspace{5 mm} The mechanical properties of matter are related to the way in which it interacts with gravitational fields \cite{Burkert:2023wzr}. The response of gravity to the presence of matter is translated into the energy-momentum tensor (EMT) such that matter mechanical properties can be extracted from its matrix elements between single particle states in terms of form factors. The general decomposition of QCD EMT in terms of gravitational form factors is given by \cite{Polyakov:2018zvc,Mamo:2019mka,Mamo:2021krl}
\begin{equation}
    \bra{p_2}T^{\mu\nu}(0)\ket{p_1}=\bar{u}(p_2)\left({A(q^2)}\,\gamma^{(\mu}P^{\nu)}+{B(q^2)}\,\frac{i\,P^{(\mu}\sigma^{\nu)\alpha}q_{\alpha}}{2\,m_N}+{C(q^2)}\,\frac{q^\mu q^\nu-\eta^{\mu\nu}q^{2}}{m_N}\right)u(p_1),\label{eq:ME}
\end{equation}
with {kinematic variables $P=\frac{1}{2}(p_2+p_1)$ and $q=(p_2-p_1)$. We also have $a^{(\mu}b^{\nu)}=\frac{1}{2}(a^\mu\,b\nu+a^\nu\,b^\mu)$},  and $\bar{u}u=2\,m_N$, with $m_N$ the nucleon mass. {We consider the usual spin operator $\sigma^{\mu\nu}=\frac{i}{2}\{\gamma^\mu,\gamma^\nu\}$.}  The scalar coefficients $A$, $B$, and $C$ are the gravitational form factors, which encode information on momentum distribution, spin and angular momentum, as well as internal mechanical forces distributions inside the nucleon, respectively.

The energy-momentum tensor can be obtained by varying the action coupled to classical gravity with respect to the metric $g^{\mu\nu}$
\begin{equation}
    T^{\mu\nu}(x)=\frac{2}{\sqrt{-g}}\frac{\delta\,S}{\delta\,g^{\mu\nu}(x)},\label{eq:Tmunu}
\end{equation}
with
\begin{equation}
    S=S_\text{G}+S_\psi,
\end{equation}
where
\begin{align}
    S_\text{G}&=\frac{1}{2\,\kappa_5^2}\int dz\,d^4x\,\sqrt{-g}\left(\mathcal{R}-2\,\Lambda\right),\\
    S_\psi&=\int d z \, d^{4}x \,\sqrt{-g}\,{\bar\Psi}({\slashed{D}}- m_{5})\Psi.
\end{align}
$\mathcal{R}$ is the Ricci scalar, $\kappa_5^2\equiv 8\pi\,G_5$, with $G_5$ Newton's constant in five dimensions, and $\Lambda$ a cosmological constant.

\begin{figure}[ht!]
    \centering
\includegraphics[width=0.9\linewidth]{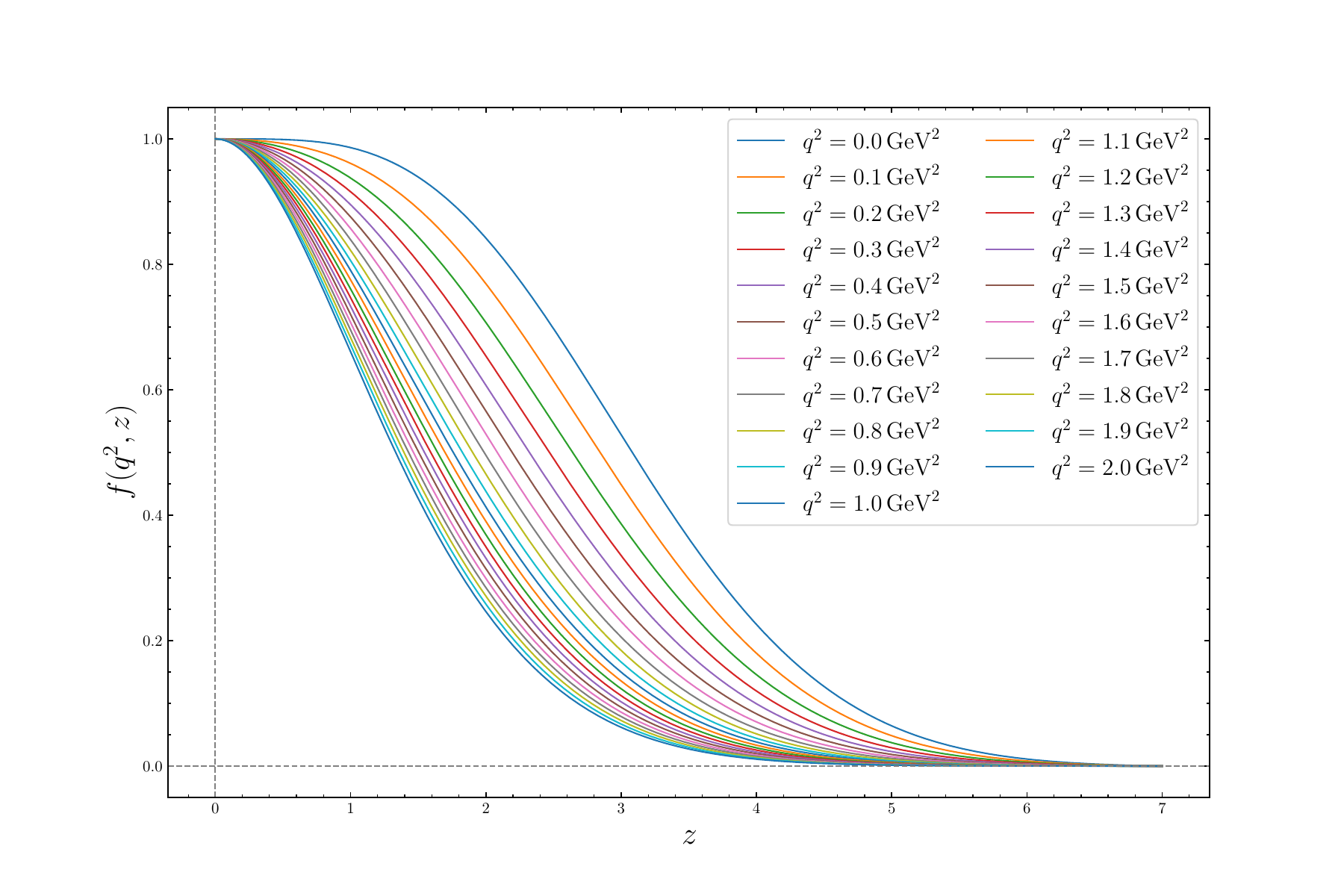}
    \caption{\sl {Graviton solutions $f(q^2,z)$, which represent both $h(q^2,z)$ and $H(q^2,z)$.}}
    \label{Fig:gravitons}.
\end{figure}

In holography, the EMT sources a fluctuation in the bulk metric such that
\begin{equation}
    g_{mn}(x,z)\to g_{mn}(x,z)+h_{mn}(x,z),
\end{equation}
with line element given by \eqref{eq:ZTmetric}, and warp factor {$\mathcal{A}(z)=-\log (z/R) +k\,z^2/2$.} The form factors are obtained by coupling the metric fluctuation $h_{mn}$ to the Dirac fermions $\psi_{L/R}$ present in the bulk EMT. To reproduce its tensorial structure in (\ref{eq:ME}), we decompose the metric fluctuation into a transverse-traceless part $h(q^2,z)$, and a longitudinal trace full part $H(q^2,z)$
\begin{equation}
    h_{mn}(x,z)=\epsilon^{TT}_{mn}\,e^{-i\,q\cdot x}\,h(q^2,z)+q_{m}\,q_n\,e^{-i\,q\cdot x}\,H(q^2,z),\label{eq:Mperturb}
\end{equation}{with a plane wave functional form for each of them}. In the axial gauge $h_{mz}={h_{zz}}=0$, the corresponding graviton equation of motion obtained from the linearized Einstein's equations takes the form of
\begin{equation}
    \frac{1}{\sqrt{-g}}\,\partial_m(\sqrt{-g}\,g^{mn}\partial_n\,h_{\mu\nu})=0,\label{eq:graviton}
\end{equation}
{Since each component in \eqref{eq:Mperturb} satisfies \eqref{eq:graviton}, we shall call them generically $f(q^2,z)$, which can be both $h(q^2,z)$ and $H(q^2,z)$. Also, by taking into account \eqref{eq:ZTmetric} along with \eqref{WF} in \eqref{eq:graviton}, we arrive at the following equation for the graviton modes}
\begin{equation}
    f^{\prime\prime}(q^2,z)+3\,\mathcal{A}^{\prime}(z)\,f^\prime(q²,z)-q^2\,f(q^2,z)=0,\label{eq:fgraviton}
\end{equation}{where prime denotes derivative with respect to the holographic coordinate, $z$. With the plane wave ansatz, we seek solutions such that $f(q^2,0)=1$ and $f(q^2, \infty)=0$. In solving numerically Eqs.~\eqref{eq:fgraviton}, we consider setting $q^2_{\text{max}}=2\,\text{GeV}^2$. This is because we want to compute form factors in the range $0\leq q^2\leq 2\,\text{GeV}^2$, which corresponds to the kinematical range used in recent lattice QCD calculations, whose results we will compare to ours in the following. In this way, we get the following set of solutions of Eqs.~\eqref{eq:fgraviton} illustrated by Fig. \ref{Fig:gravitons}.} 
{ With these solutions at hand, we can proceed to compute the the gravitational form factors. One can start by considering \eqref{Action} as the matter bulk action,}
\begin{equation}
    S_\psi=\int dz\,d^4x\, e^{5\mathcal{A}(z)}\left[\bar{\Psi}\left(e^{-\mathcal{A}(z)}\gamma^5\,\partial_z+e^{-\mathcal{A}(z)}\,\gamma^\mu\partial_\mu+2\,\mathcal{A}^\prime(z)\,\gamma^5-m_5\right)\Psi\right].\label{eq:gaction}
\end{equation}
{This enters \eqref{eq:Tmunu} as $S_\psi$, which will give rive to the form factors}. As a consequence of this choice, one can show that there will be no coupling between metric fluctuation and a spin term $\sigma^{\mu}$ in the deformed AdS model, resulting in a vanishing $B$ form factor.

%%%%%%%%%%%%%%%%%%%%%%%%%
%%%%%  Subection  %%%%%%%
%%%%%%%%%%%%%%%%%%%%%%%%%

\subsection{$A(q^2)$ form factor}
\label{AFF}

The $A(q^2)$ form factor is obtained by contracting (\ref{eq:ME}) with $\epsilon_{\mu\nu}^{TT}$
\begin{equation}
    \bra{p_2}\epsilon_{\mu\nu}^{TT}\,T^{\mu\nu}\ket{p_1}=\bar{u}(p_2)\,A(q)\,\epsilon_{\mu\nu}^{TT}\,{\gamma^\mu} p^\nu\,u(p_1),\label{Acontraction}
\end{equation}
where
\begin{equation}
    A(q^2)=\frac{1}{2}\int\,dz\,e^{-5\,\mathcal{A}(z)}(\psi_L^2+\psi_R^2)\,h(q^2,z).\label{eq:AFF}
\end{equation}

Using the numerical solutions of \eqref{SchEq} and \eqref{eq:fgraviton}, we numerically solve  the integral in \eqref{eq:AFF} as a function of $q^2$. In order to compare it with the lattice QCD results, we consider a normalization of our numerical $A(q^2)$ by $A(0)$ of the lattice data, since this initial value is not fixed in holography. Using an analytical soft-wall model, the authors in \cite{Mamo:2019mka} proceeded by choosing $A(0)$ from the lattice data of \cite{Shanahan:2018pib} to fix their corresponding zero momentum squared value and adjust their analytical result to these data. This can be achieved by considering a target value, say $A(0)_\text{targ}$ from the lattice, and then a normalized curve given by
\begin{equation}\label{normproc}
    A(q^2)_\text{norm}=A(0)_\text{targ}\,\frac{A(q^2)}{A(0)},
\end{equation}
{such that $A(0)_\text{norm}$ coincides with the target value from the lattice.}

\begin{figure}[ht!]
    \centering
\includegraphics[width=0.8\linewidth]{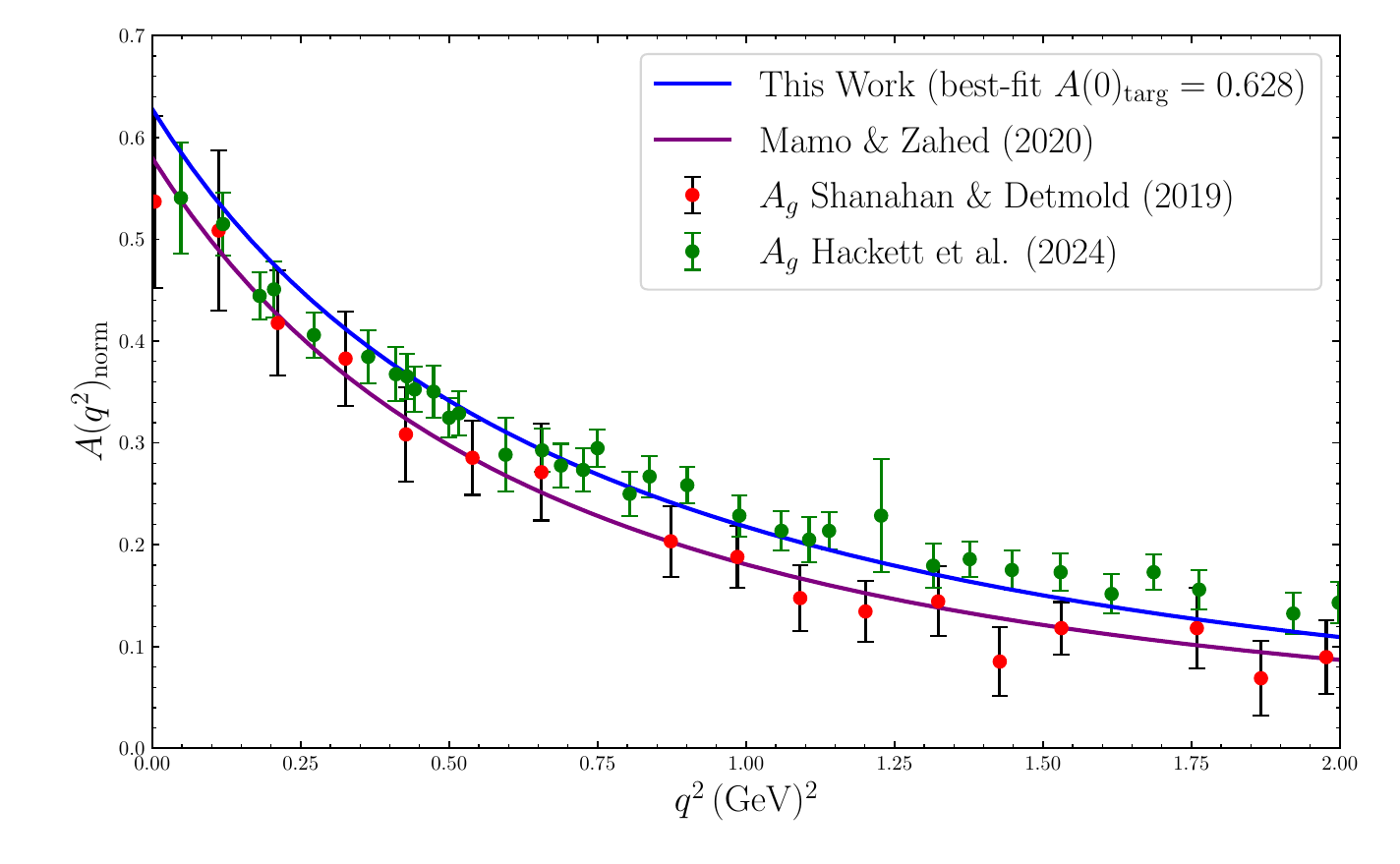}
    \caption{\sl $A(q^2)_\text{norm}$ form factor adjusted to the gluon contributions of lattice data of \cite{Shanahan:2018pib} and \cite{Hackett:2023rif}. The green dots correspond to the gluonic contribution in lattice QCD of \cite{Hackett:2023rif}, and the blue curve represents our best-fit normalized holographic form factor $A$ .The purple curve represents the soft-wall result of Mamo \& Zahed \cite{Mamo:2019mka}, compared to the lattice data of \cite{Shanahan:2018pib} (red dots).}
    \label{Fig:AFit}.
\end{figure}

Here we choose to compare our holographic model with the more up to date Lattice data of \cite{Hackett:2023rif} related to the gluon contribution to $A(q^2)$. In this work, however, we extend the normalization procedure described above by choosing a range of target $A(0)_\text{targ}$ values that comprises the zero momentum values of \cite{Shanahan:2018pib} and \cite{Hackett:2023rif}, as well as their error bars. We choose this range as $A(0)_\text{targ}=0.4- 0.7$. Then, we perform a $\chi^2$ minimization procedure as follows. The following $\chi^2$ functional form is given by
\begin{equation}\label{chi2}
    \chi^2=\sum_{i=1}^N\frac{[A(q^2_i)_\text{norm}-A(q^2_i)_\text{lat})]^2}{\sigma^2},
\end{equation}
$N$ represents the total number of momenta transferred.  $\sigma=0.5\,(A_\text{upper}-A_\text{lower})$ corresponds to the symmetric error associated with the upper and lower points of the error bars of the data from \cite{Hackett:2023rif}. Keeping all other parameters in the model fixed, we look for the value of $A(0)_\text{targ}$ in $A(q^2_i)_\text{norm}$ that minimizes \eqref{chi2}. In this context, we find $A(0)_\text{targ}=0.628$ to be the best target value found. We point out that this target value was entered as the only degree of freedom used in this minimization procedure. This gives $\chi^2_\text{min}=32.471$ and $\chi^2/\text{ndf}=0.984$. These results give the best-fit curve of our normalized holographic model, as illustrated in Fig. \ref{Fig:AFit}.

The gluonic radius related to the momentum distribution is given by \cite{Polyakov:2018zvc}
\begin{equation}
    \sqrt{\langle r^2_{g}\rangle}=\sqrt{-6\,\left(\frac{d\ln{A(q^2)_\text{norm}}}{d\,q^2}\right)\Bigg|_{q=0}}=0.59 \,\text{fm}.
\end{equation}

%%%%%%%%%%%%%%%%%%%%%%%%%%
%%%%%  Subsection  %%%%%%%
%%%%%%%%%%%%%%%%%%%%%%%%%%

\subsection{$D(q^2)$ form factor}
\label{DFF}

\begin{figure}[h]
    \centering
\includegraphics[width=0.8\linewidth]{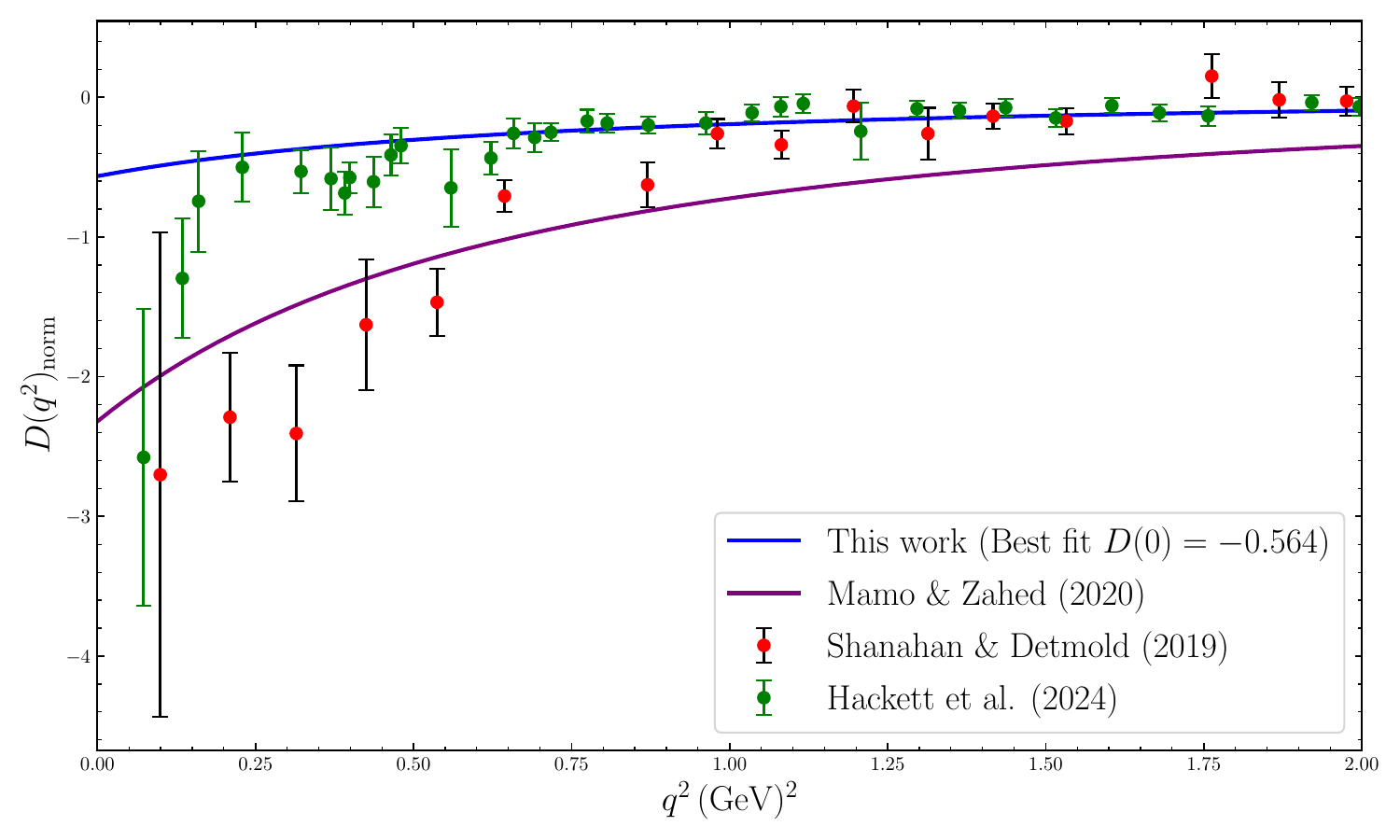}
    \caption{\sl Holographic $D(q^2)_\text{norm}$ form factor from deformed AdS model (blue curve) compared to gluon lattice data of \cite{Hackett:2023rif} (green dots). The purple curve corresponds to the soft-wall model of \cite{Mamo:2019mka}, which compares to the lattice data of \cite{Shanahan:2018pib} (red dots).}
    \label{Fig:DFit}
\end{figure}

To obtain $C$, we must contract \eqref{eq:ME} with $\eta_{\mu\nu}$, in such a way as to extract the term proportional to $q^2$, resulting in
\begin{equation}
    C(q^2)= \frac{1}{2}\int\,dz\, e^{-5\,\mathcal{A}(z)}\,H(q^2,z)\,\psi_L\,\psi_R.\label{eq:CFF}
\end{equation}
From the numerical integration of \eqref{eq:CFF}, one can calculate the form factor $D$ defined by $D(q^2)\equiv 4\,C(q^2)$. We perform the same normalization and $\chi^2$ minimization procedure for $D(q^2)$. Similarly to $A(0)_\text{targ}$, we choose to restrict the target $D(0)$ in a normalized $D(q^2)_\text{norm}$ to the range $D(0)_\text{targ}=(-1.5)-0.0$, in which most of the lattice data from \cite{Hackett:2023rif} lie. The resulting best target value $D(0)_\text{targ}$ is found to be $D(0)_\text{targ}=-0.564$. The resulting fit to the lattice data of \cite{Hackett:2023rif} is illustrated in Fig.~\ref{Fig:DFit}. We can notice that we get close to the QCD data of the lattice, but the fit is not as good as that achieved for the form factor $A$, especially for the low momenta transferred $q^2$. Indeed, the minimization procedure described above, when applied here to $D(q^2)_\text{norm}$ with a target $D(0)_\text{targ}$, gives $\chi^2_\text{min}=42.491$ with $\chi^2/\text{ndf}=1.328$, which amounts the deviation from the data
considered in Ref.\cite{Hackett:2023rif} in low q2 values. However, even though there is this discrepancy between our holographic model and the data considered for comparison, we can still explore the fact that its trend qualitatively follows the gluon lattice data of \cite{Shanahan:2018pib} and \cite{Hackett:2023rif} to study the mechanical properties of the proton, such as the pressure and shear distributions in this model, which we are going to do next.

%%%%%%%%%%%%%%%%%%%%%%%%%
%%%%%%  Section  %%%%%%%%
%%%%%%%%%%%%%%%%%%%%%%%%%

\section{Pressure, Shear, and Energy Distributions}
\label{PressureShear}

\begin{figure}[ht!]
    \centering
\includegraphics[width=0.8\linewidth]{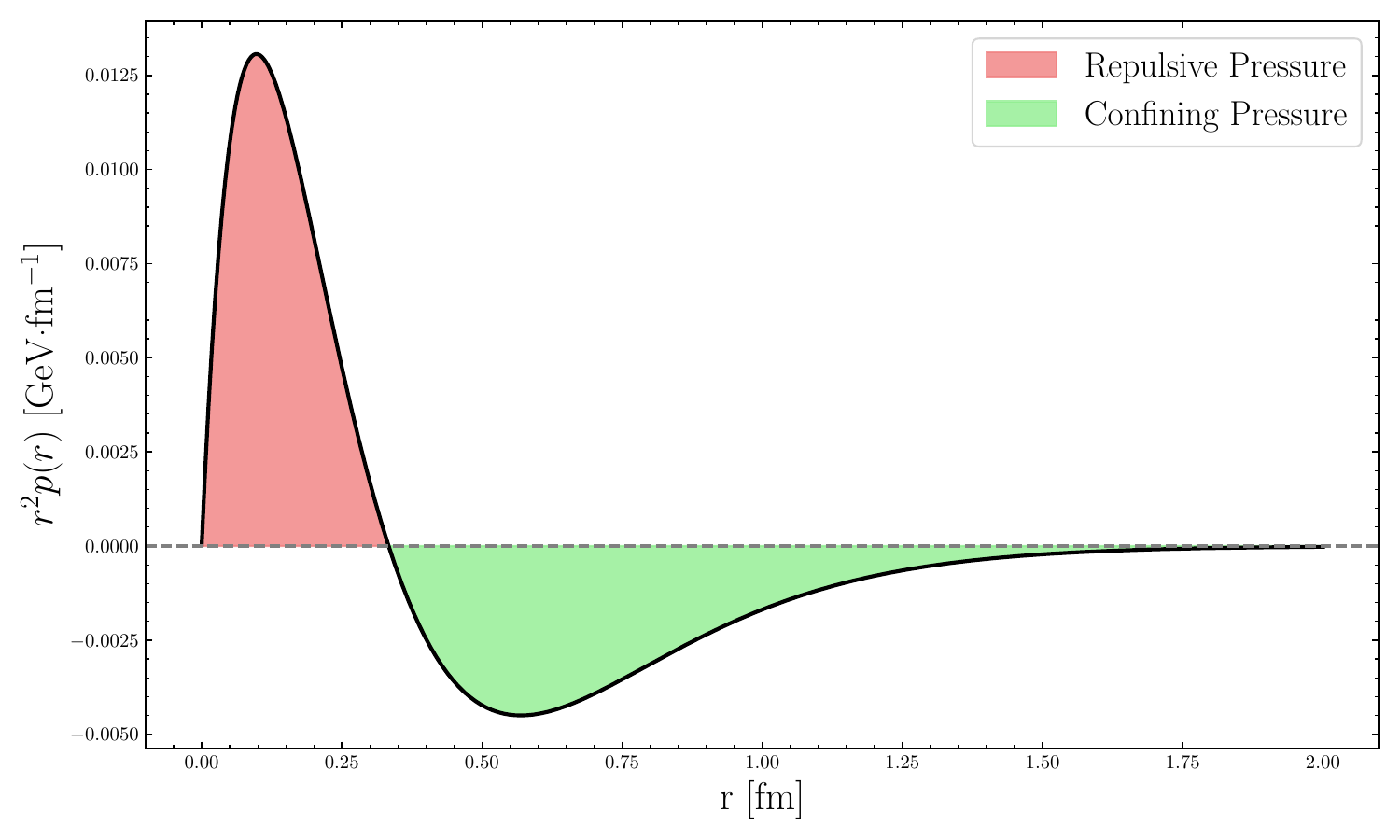}
    \caption{\sl Pressure distribution from the dipole approximation \eqref{eq:DipoleA}, with a repulsive (red) and confining (green) pressure regions.}
    \label{Fig:Pressure}
    \includegraphics[width=0.8\linewidth]{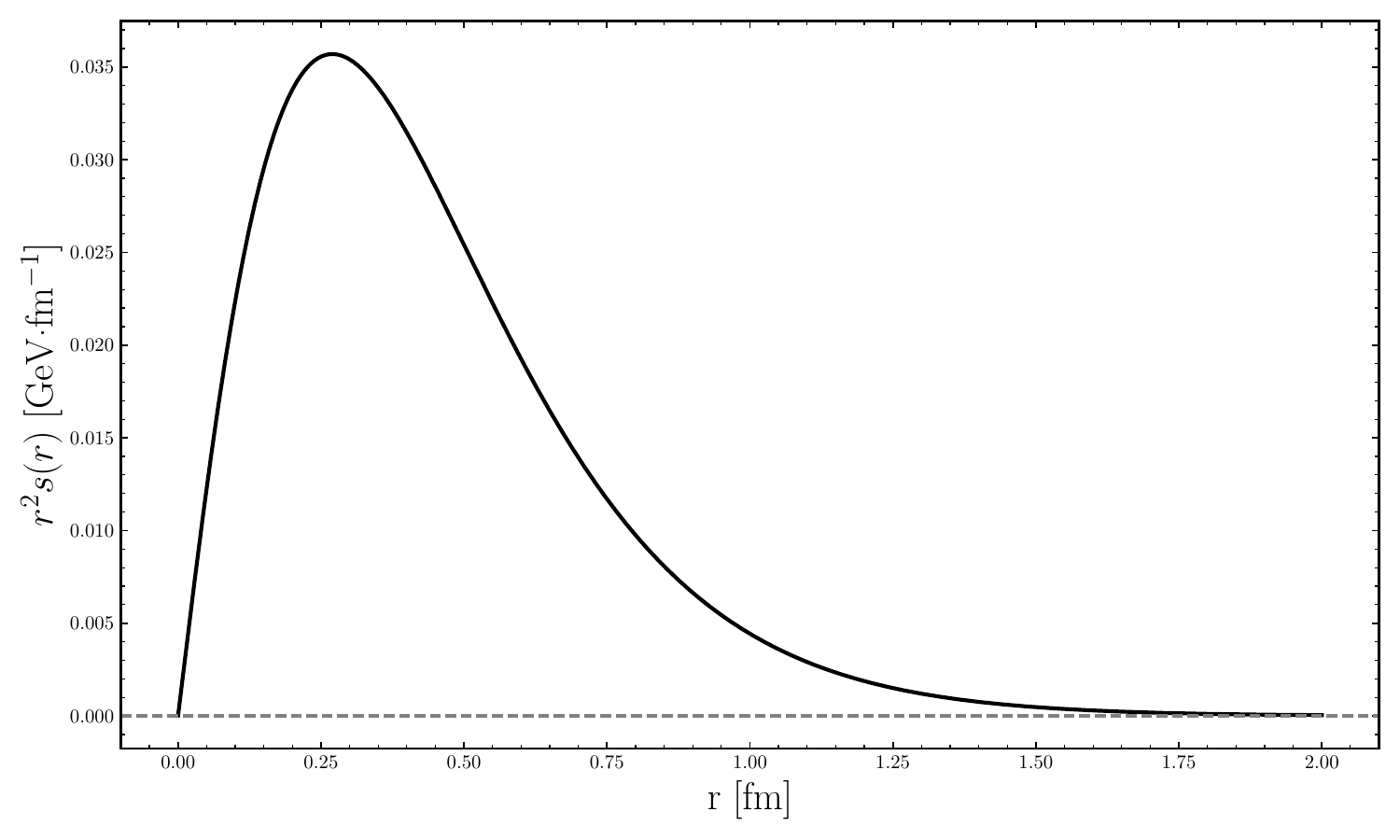}
    \caption{\sl Shear distribution from the dipole approximation \eqref{eq:DipoleA}.}
    \label{Fig:Shear}
\end{figure}

 Pressure and shear distributions in the system can be probed by Fourier transforming $D(q^2)$ to position space
\begin{equation}\label{DqFT}
    \Tilde{D}(r)=\int \frac{d^3q}{2\,m\,(2\pi)^3}e^{-i\,q\,\cdot\,r}D(q^2)_\text{norm}.
\end{equation}
This is because these distributions can be formulated in terms of $\Tilde{D}(r)$ according to \cite{Polyakov:2018zvc}
\begin{align}
    p(r)&=\frac{1}{6\,m}\frac{1}{r^2}\frac{d}{d\,r}\left(r^ 2\frac{d}{d\,r}\Tilde{D}(r)\right),\label{eq:Pressure}\\
    s(r)&=-\frac{r}{4\,m}\frac{d}{d\,r}\left(\frac{1}{r}\frac{d}{d\,r}\Tilde{D}(r)\right),\label{eq:shear}
\end{align}
with $m=M_p$ the proton mass. For numerical convenience, instead of using the numerical result for $D(q^2)_\text{norm}$, we use a dipole approximation of it, which is well behaved when computing its Fourier transform.  We consider
\begin{equation}
    D(q^2)=\frac{D_0}{\left(1+\frac{q^2}{\mathcal{E}^2}\right)^2},\label{eq:DipoleA}
\end{equation}
where $D_0$ and $\mathcal{E}$ are free parameters. A very good fit to our numerical result $D(q^2)_\text{norm}$ is found with $D_0=-0.556$ and $\mathcal{E}=1.184\,\text{GeV}$. The pressure and shear distributions are shown by Figures \ref{Fig:Pressure} and \ref{Fig:Shear}. 

One can also obtain the mechanical radius associated to those distributions, given by \cite{Polyakov:2018zvc}
\begin{equation}
    \sqrt{\langle\,r_{\text{mech}}^2\rangle}=\sqrt{\frac{6}{\mathcal{E}^2}}=0.41\,\text{fm}.
\end{equation}
These results are comparable with experimental results in \cite{Burkert:2018bqq} and lattice QCD ones, as in \cite{Shanahan:2018nnv}. 

One can notice from these predictions that the strong force behaves as a repulsive force in the inner regions of the proton, whereas it behaves as a confining force in its outer regions. These positive and negative pressure regions present in Fig. \ref{Fig:Pressure} integrate to $9.19\times 10^{-7}\,\text{GeV}$, which, within the precision of our calculation, is numerically consistent  with the von Laue stability condition \cite{laue1911dynamik}
\begin{equation}
    \int_ 0^\infty r^2\,p(r)\,dr=0,
\end{equation}
that describes how internal pressure forces balance inside a composed particle, such as a proton. The shear distribution follows the pressure along the way, increasing as the pressure is repulsive and decreasing as it becomes more confining, acting sideways to keep the system balanced and stable.  This picture can be further supported by exploring tangential and normal forces in the proton, which we are going to explore in the next section.

\begin{figure}[ht!]
    \centering
    \includegraphics[width=0.9\linewidth]{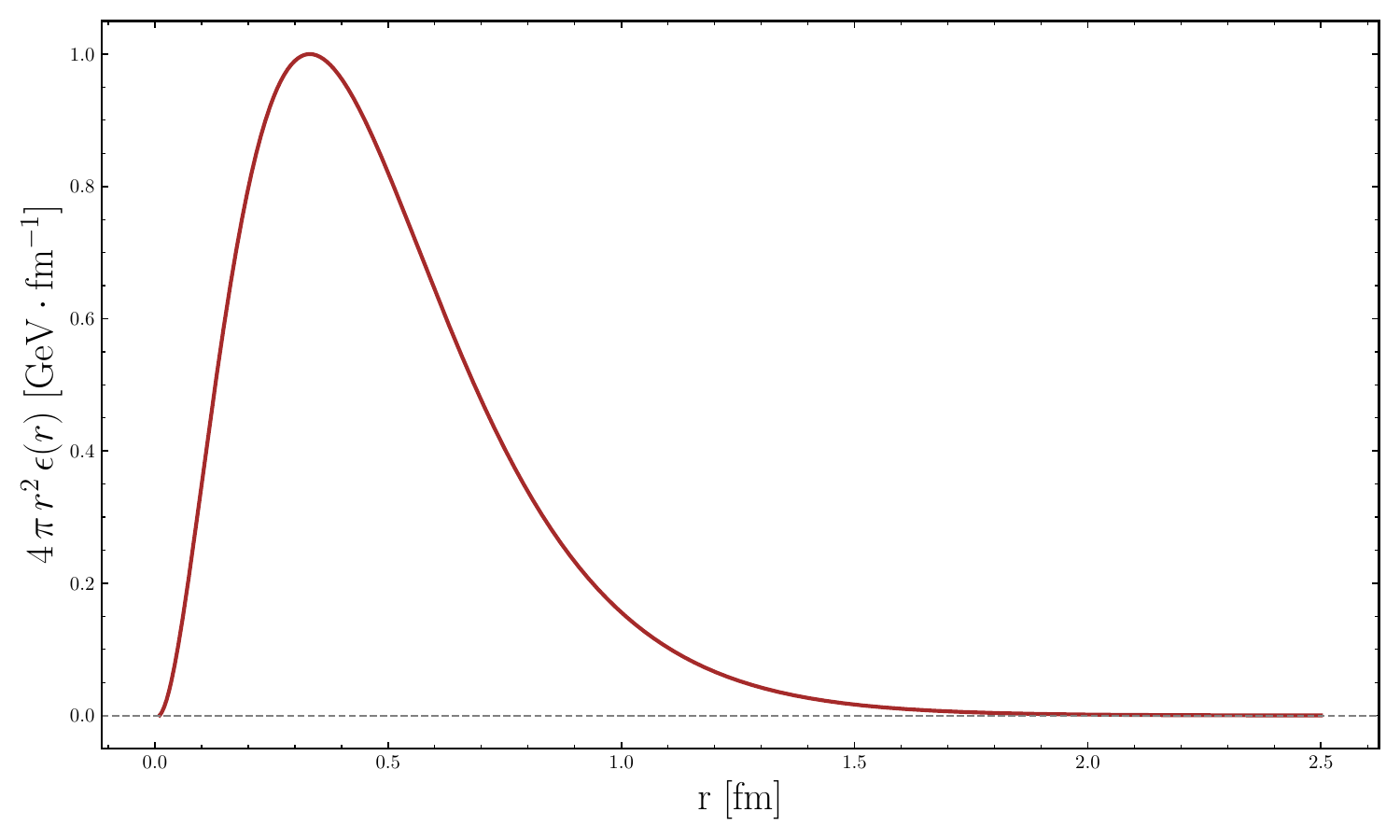}
    \caption{\sl Energy distribution from the dipole approximations \eqref{eq:DipoleA} and \eqref{eq:Adipole}.}
    \label{Fig:Energy}
\end{figure}

The energy distribution in the system is obtained by
\begin{equation}\label{eq:energy}
    \epsilon(r)=m\,\left[A(q^2)-\frac{q^2(D(q^2)+A(q^2)-2\,B(q^2))}{4\,m^2}\right]_{\text{FT}},
\end{equation}
where FT stands for Fourier transform. Here, the $A$ and $D$ form factors represent our normalized ones obtained in Sect. \ref{GFFs} and, in our case, $B=0$. We also need to obtain an approximation to our numerical $A(q^2)_\text{norm}$ to compute the Fourier transform above, and again consider a dipole parameterization of the form
\begin{equation}\label{eq:Adipole}
    A(q^2)=\frac{A_0}{\left(1+\frac{q^2}{\lambda^2}\right)},
\end{equation}
with $A_0$ and $\lambda$ free parameters. The best approximation to our numerical $A(q^2)_\text{norm}$ is found with
$A_0=0.614$ and $\lambda=1.201\,\text{GeV}$. Hence, the resulting energy density can be computed and is displayed in Fig. \ref{Fig:Energy}.

%%%%%%%%%%%%%%%%%%%%%%%%%
%%%%%%  Section  %%%%%%%%
%%%%%%%%%%%%%%%%%%%%%%%%%

\section{Normal and Tangential Forces}
\label{Forces}

\begin{figure}[ht!]
    \centering
\includegraphics[width=0.8\linewidth]{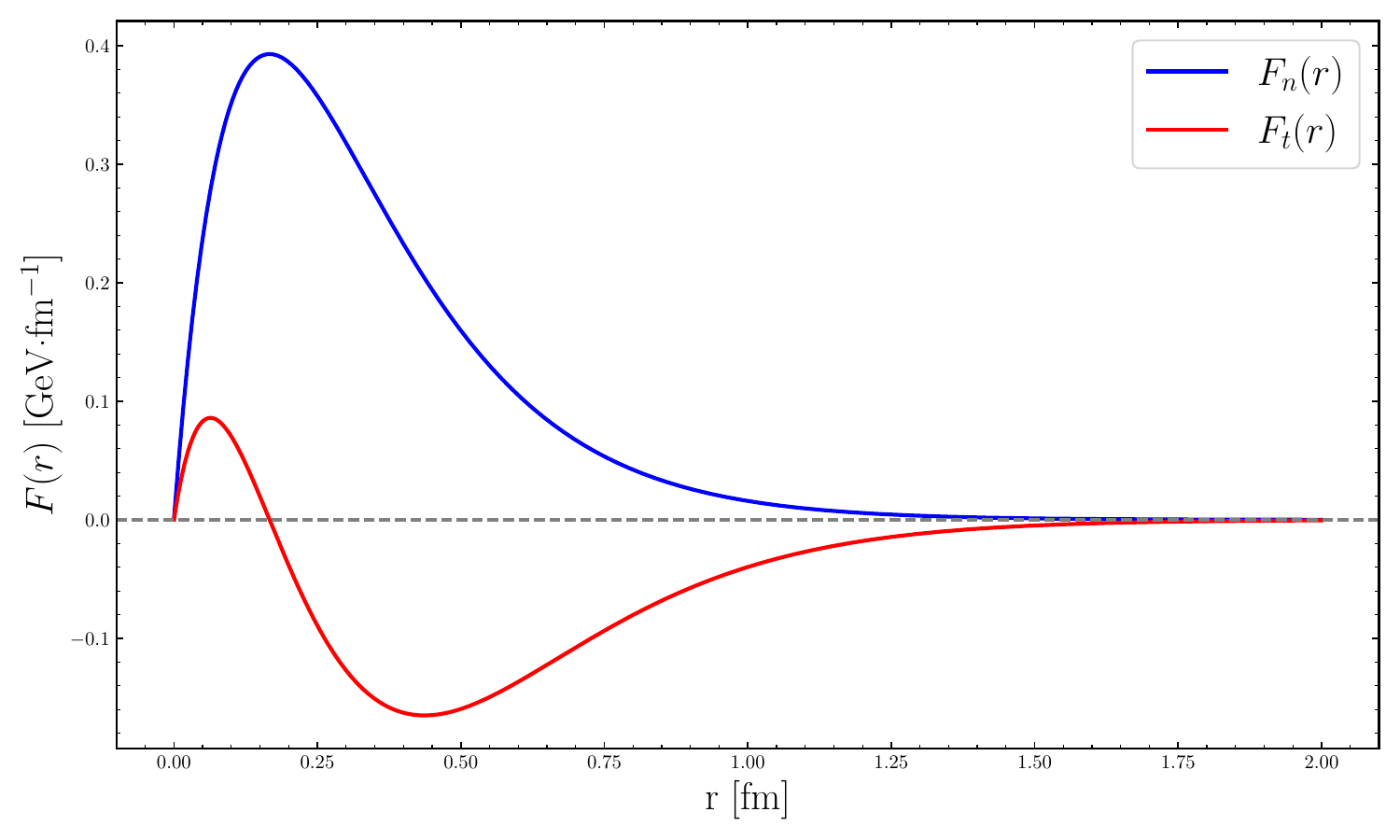}
    \caption{\sl Normal and tangential forces on a spherical shell of radius $r$ in the nucleon.}
    \label{Fig:Forces}
\end{figure}

\begin{figure}[htbp]
    \centering
    % First subfigure
    \begin{subfigure}[b]{0.45\textwidth}
        \centering
        \includegraphics[width=\textwidth]{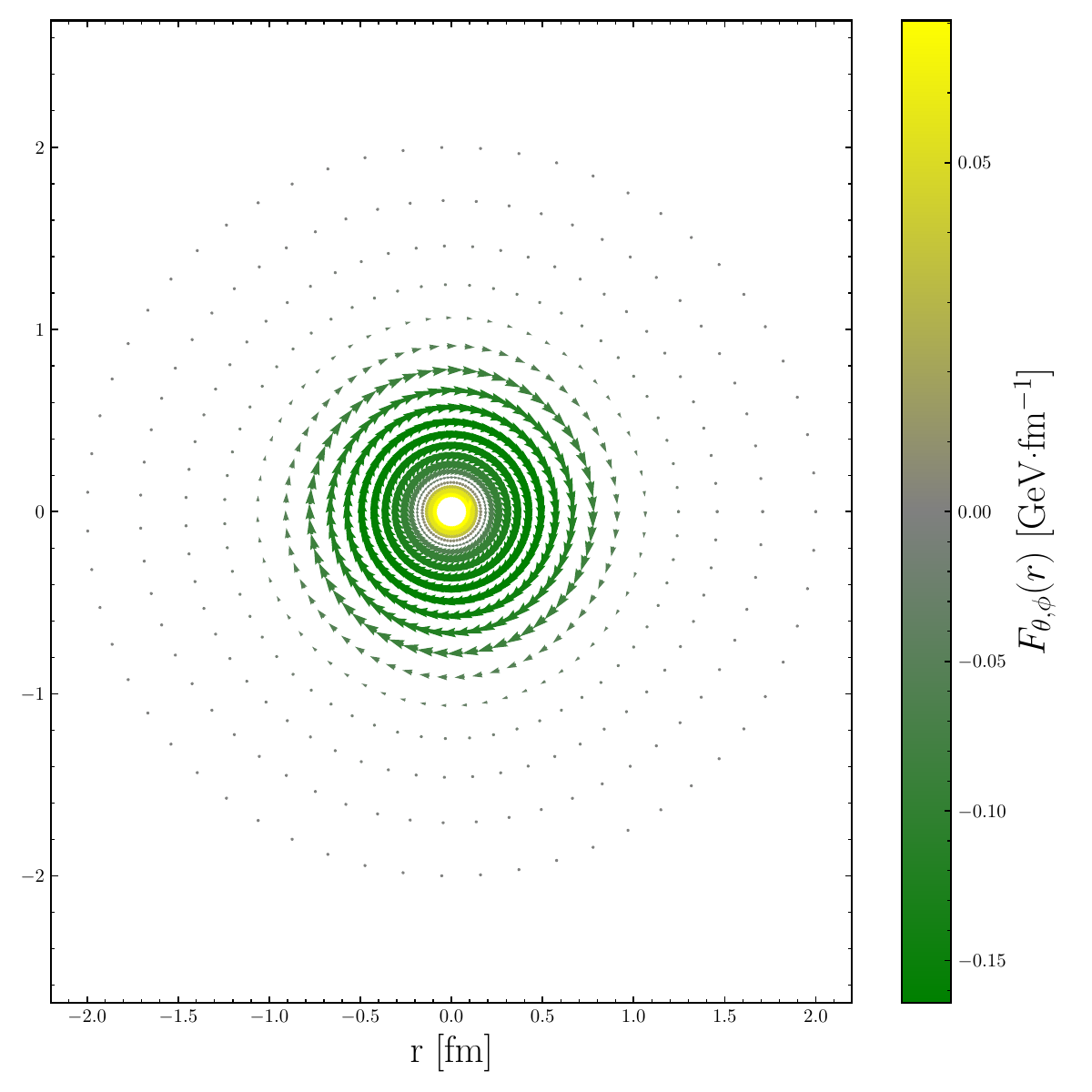}
        \label{fig:first}
    \end{subfigure}
    \hfill
    % Second subfigure
    \begin{subfigure}[b]{0.45\textwidth}
        \centering
        \includegraphics[width=\textwidth]{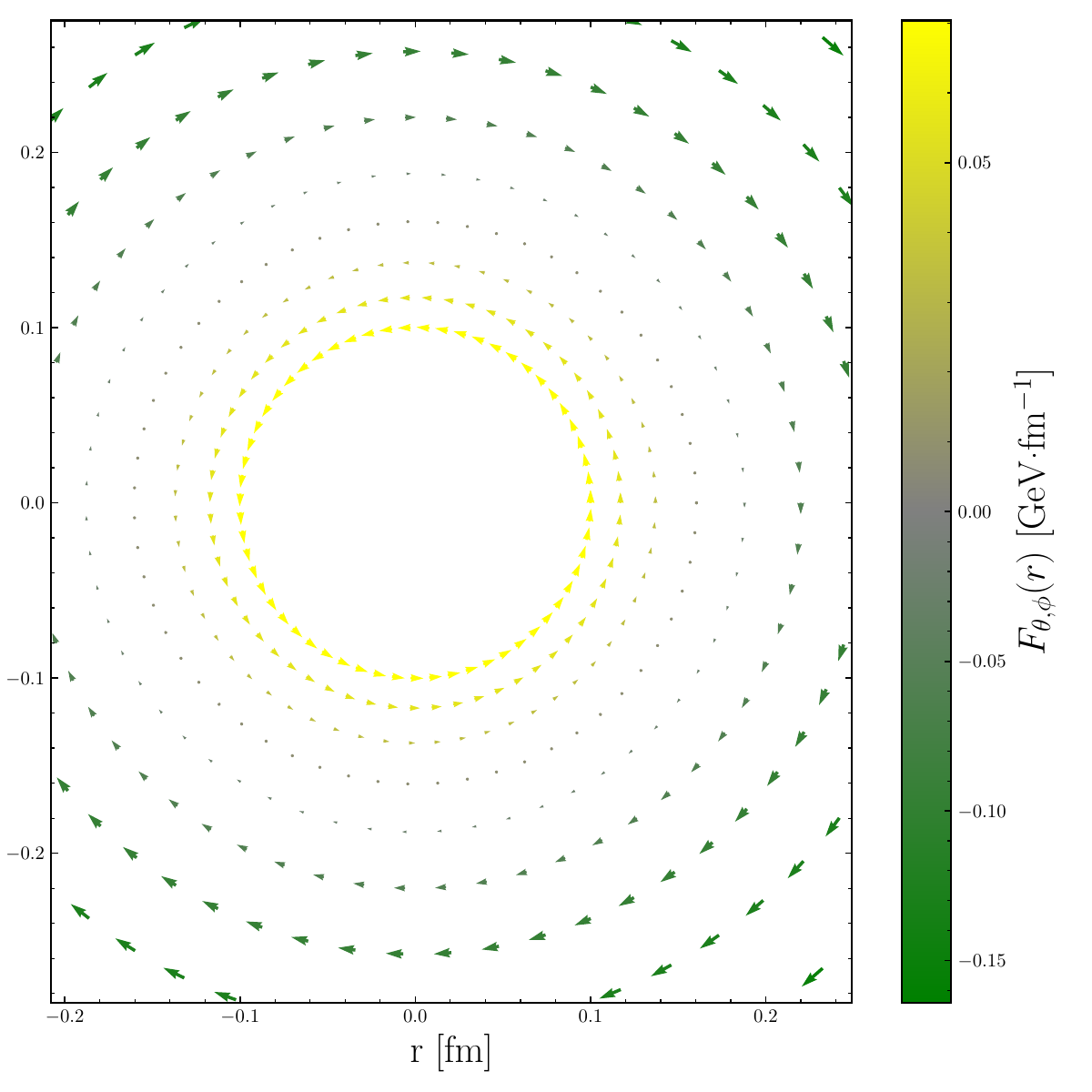}
        \label{fig:second}
    \end{subfigure}

    \caption{Force field visualization of the tangential forces $F_\theta$ and $F_\phi$ as functions of $r$. The left panel displays the full profile of these forces. The right panel zooms in into the inner region where the sign of the forces changes.}
    \label{fig:side_by_side}
\end{figure}

\begin{figure}[ht!]
    \centering
\includegraphics[width=0.8\linewidth]{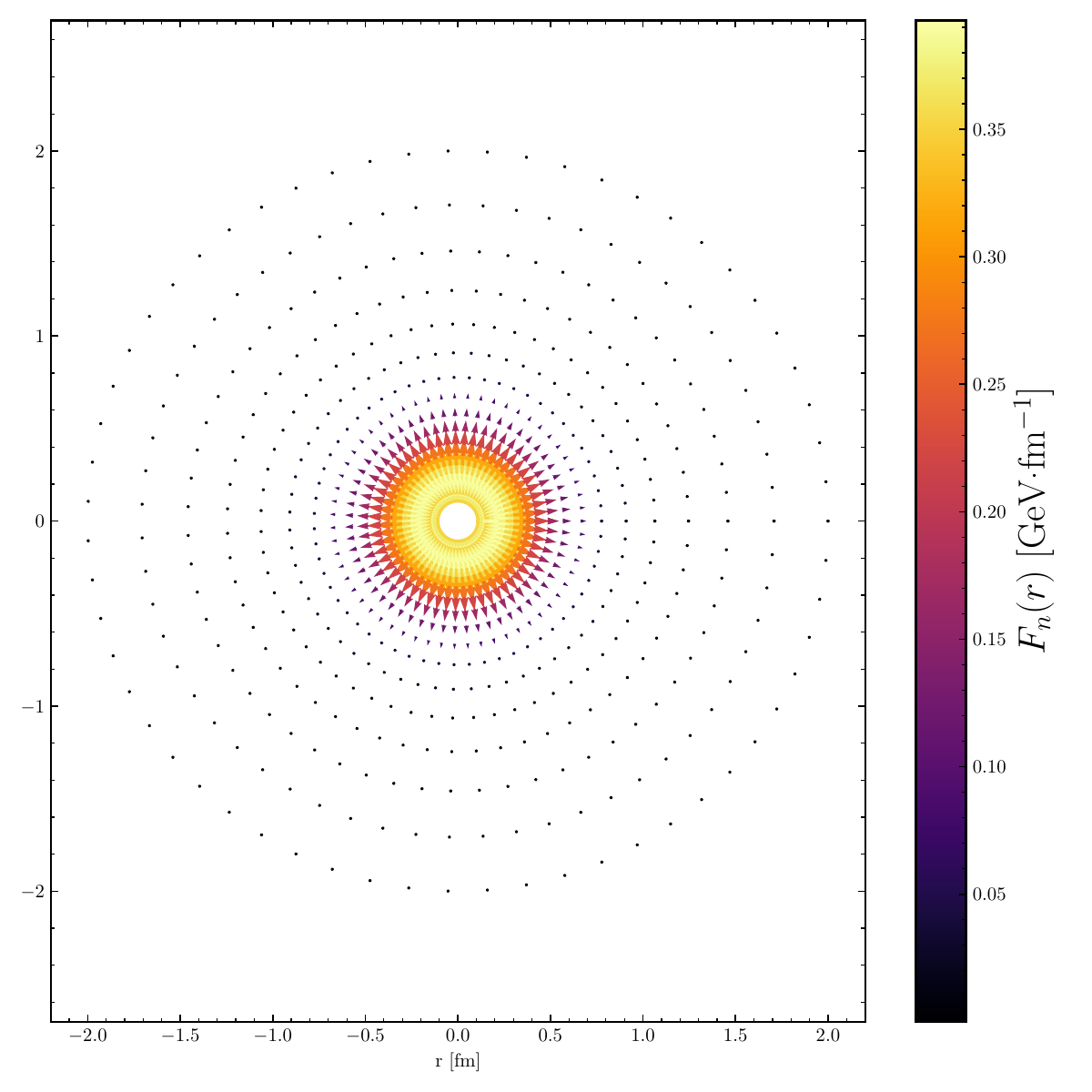}
    \caption{\sl Force field visualization of the normal force $F_n(r)$ inside the proton.}
    \label{Fig:Normal}
\end{figure}

\hspace{5mm}Following \cite{Polyakov:2018zvc}, we consider a slice of the proton defined in spherical coordinates by $\theta=\pi/2$. The forces acting on the infinitesimal area element $d\textbf{S}=dS_n\,\textbf{e}_n+dS_\theta\,\textbf{e}_\theta+dS_\phi\,\textbf{e}_\phi+$ of the slice are
\begin{equation}\label{infitesimalforces}
    \frac{dF_n}{dS_n}=\frac{2}{3}s(r)+p(r),\quad\frac{dF_\theta}{dS_\theta}=\frac{dF_\phi}{dS_\phi}=-\frac{1}{3}s(r)+p(r),
\end{equation}
with the tangential forces being equal when considering the proton as a spherical symmetrical system. Then, the normal and tangential forces on a spherical shell of radius r in the proton are $F_n$ and $F_t$,
\begin{equation}\label{forces}
    F_n=4\,\pi\,r^2\left[\frac{2}{3}s(r)+p(r)\right],\quad\,F_t=4\,\pi\,r^2\left[-\frac{1}{3}s(r)+p(r)\right],
\end{equation}
which are displayed in Fig.~\ref{Fig:Forces}.
Regarding the normal force, its positive values correspond to stretching in the system, which is present mainly in the inner regions of the proton. The change in sign of the tangential forces acts to counterbalance any possible squeeze from the confining pressure so that it averages to zero. This behavior of the tangential forces can also be seen from the force field point of view, as illustrated by Fig.~\ref{fig:side_by_side}. A force field visualization for $F_n(r)$ is also displayed in Fig. \ref{Fig:Normal}.

%%%%%%%%%%%%%%%%%%%%%%%%%
%%%%%%  Section  %%%%%%%%
%%%%%%%%%%%%%%%%%%%%%%%%%

\section{Conclusions}
\label{conclusions}

\hspace{5mm} 
In the present work, we employ the deformed AdS background holographic model to describe gravitational form factors of the proton and explore some of its mechanical properties based on the $D$ term, focusing on the contribution of the gluon to them. We find a very good agreement between our prediction for the gluonic form factor $A$ and the recent lattice QCD computation performed in \cite{Hackett:2023rif}, using a normalization procedure to fit $A(0)$ to the lattice result with a target $A(0)_\text{targ}$ that minimizes the $\chi^2$ measure of the difference between our model prediction and the lattice data. Its trend is also compared to the previous holographic result of \cite{Mamo:2019mka} as well as to the previous lattice data of \cite{Shanahan:2018pib}. The same approach is carried out for the gluonic term $D$, resulting, in this case, in a qualitative agreement between our model and the lattice data of \cite{Hackett:2023rif}, whose behavior draws a particular shape not fully captured by the holographic model used in this work. However, we remark that those results were obtained numerically in all the steps present in their calculations, in contrast to previous holographic results using the soft-wall model in \cite{Abidin:2009hr,Mamo:2019mka}, which use analytical solutions to form factors. The general agreement between the numerical results in our work, which are also in the ballpark of the results present in \cite{Shanahan:2018pib,Mamo:2019mka}, and the lattice results of \cite{Hackett:2023rif}, shows that this model is capable of capturing features present in both the lattice QCD literature and previous holographic models.

From the $D$ term, since its trend follows the behavior present in the lattice results, we can explore how the model describes some mechanical properties of the nucleon. Using a dipole approximation to our numerical result, which is better suited for the calculation of the Fourier transform of $D(r)$ and $A(r)$, we can compute the pressure and shear distributions inside the proton, which balance each other to keep the system stable. This is supported by the small value of the integral $\int r^2\,p(r)\,dr=9.19\times10^{-7}\,\text{GeV}$ in our approximation, which should be zero in the exact case. We are also able to evaluate the energy density in the proton using a dipole approximation to $A(q^2)_\text{norm}$. The results follow similar trends present in the literature, e.g., such as those in \cite{Lorce:2018egm,Polyakov:2018zvc,Shanahan:2018nnv,Burkert:2018bqq,Burkert:2021ith,Mamo:2019mka,Mamo:2021krl}. However, the mechanical radius associated with these distributions is smaller than the ones found in these works. From pressure and shear distributions, internal forces are also investigated, presenting a similar behavior appearing in this literature.

{ We remark that even though the calculations performed in this work follow a similar receipt in holographic modeling as the ones performed in \cite{Abidin:2009hr,Mamo:2019mka} and have comparable results, they are obtained in a fundamentally different manner on theoretical grounds. The approach used here within the deformed AdS model considers a direct deformation of the anti-de Sitter space by a dilaton-like term \eqref{WF} present in its metric \eqref{eq:ZTmetric}, as opposed to the soft-wall model we briefly discussed in Appendix \ref{SoftWall}. This simple change makes a real difference, especially in the description of fermions. Beyond fermions, the graviton solutions in the soft-wall model are analytical, while in the deformed AdS model they are numerical, which consequently leads to numerical form factors as the ones shown in Sect. \ref{GFFs}. This casts the whole approach using the deformed AdS model fundamentally different from the one using the soft-wall model, resulting in a new prediction of the gravitational form factors  parameterizing the proton matrix elements of the QCD energy-momentum tensor, as well as the resulting mechanical properties associated with it.}

We plan to further extend these discussions to the case of the pion in the present holographic model in a future work, since this hadron's GFFs were less explored in previous holographic models, and lattice data are also available to be compared with. We hope that this work can help future experiments to be carried out by the EIC in the near future, by supporting the predictions to the internal forces acting inside the proton and how the strong nuclear force manifests in hadronic systems.

%%%%%%%%%%%%%%%%%%%%%%%%%%%%%%%%%%%%%%%%%%
%%%%%%%%%%%%%%%%%%%%%%%%%%%%%%%%%%%%%%%%%%
%%%%%%%%%%%%%%%%%%%%%%%%%%%%%%%%%%%%%%%%%%

\section*{Data Availability Statement} The data that support the findings of this article were
extracted from Refs.~\cite{Shanahan:2018pib,Hackett:2023rif} and the numerical implementation used in this work is openly available at Ref.~\cite{nascimento_2026_19713524}.

%%%%%%%%%%%%%%%%%%%%%%%%%%%%%%%%%%%%%%%%%%
%%%%%%%%%%%%%%%%%%%%%%%%%%%%%%%%%%%%%%%%%%
%%%%%%%%%%%%%%%%%%%%%%%%%%%%%%%%%%%%%%%%%%

\section*{Acknowledgments}
We would like to thank Jorge Noronha for enlightening discussions. This work is supported by Coordenação de Aperfeiçoamento de Pessoal de Nível Superior (CAPES), under finance code 0001. A. N. acknowledges the support by the Brazilian National Council for Scientific and Technological Development (CNPq) under grant No. 176343/2025-3. HBF is partially supported by Conselho Nacional de Desenvolvimento Cient\'{\i}fico e Tecnol\'{o}gico (CNPq) under grant  310346/2023-1, and Fundação Carlos Chagas Filho de Amparo à Pesquisa do Estado do Rio de Janeiro (FAPERJ) under grant E-26/204.095/2024.

%%%%%%% Brief Review on the Soft-Wall Model %%%%
{
\appendix
\section{Brief Review of the Soft-Wall Model for Fermions}
\label{SoftWall}
\hspace{5mm}Here we present a brief review of the soft-wall model, one of the first holographic models used in the phenomenology of QCD. It's action is given by \cite{Karch:2006pv,Colangelo:2008us}
\begin{equation}
    S=\int d^5{x}\sqrt{-g}\,e^{-\kappa^{2}\,z^{2}}\,\mathcal{L},\label{SWact}
\end{equation}
with $\kappa$ having dimension of mass and the exponential factor known as the \textit{dilaton}, $\mathcal{L}$ is the five-dimenisonal Lagrangian of the relevant fields of the model and $g$ is determinant of the metric of the anti-de Sitter space with line element
\begin{equation}
    ds^2=g_{mn}\,dx^mdx^n=\frac{R^2}{z^2}(dz^2+\eta_{\mu\nu}\,dy^\mu dy^\nu),\label{eq:AdS}
\end{equation}
where $R$ is the curvature radius of the AdS spacetime. The indices $\{m,n\}$ refer to the five-dimensional space, $\{\mu,\nu\}$ to the four dimensional Minkowski ones representing the boundary space, and $z$ is the holographic coordinate. The dilaton factor plays the role of a ``soft wall" in the equations of motion that can be derived from \eqref{SWact}, limiting the edges of the anti-de Sitter space, breaking the conformal symmetry of its geometry. 
%\subsection{The particular case of fermions}
For fermions, a special version of this model is necessary in order to have a discrete mass spectrum. This is because in the case of a Dirac action, the exponential factor that carries the dilaton factors out of the equations of motion. In this way, the fermion dynamics is not affected at all by such a term. A way to fix this issue is to add a term depending on the $z$ coordinate in the mass term of the action as an effective potential as in the following manner \cite{Abidin:2009hr,Mamo:2019mka} 
\begin{equation}
    S=\int d^{5}x\,e^{-\kappa^2z^2} \,\sqrt{-g}\,{\bar\Psi}\left[{\slashed{D}}- \left(m_{5}+\frac{\kappa^2z^2}{R^2}\right)\right]\Psi,\label{MSWAction}
\end{equation}
with $m_5$ the fermionic bulk mass in five dimensions. Here, we see how the original model is generally adjusted to describe fermions. Explicitly, one can write \cite{Abidin:2009hr, Braga:2011wa}):
\begin{equation}\label{FermionActionDSW}
    S=\int dz\,d^{4}x\,\sqrt{-g}\,e^{-\Phi(z)}\,\left[\frac{i}{2}\bar\Psi\,e^{m}_{a}\,\Gamma^{a}\,\mathcal{D}_{m}\Psi-\frac{i}{2}(\mathcal{D}_{m}\,\Psi)^{\dagger}\,\Gamma^{0}\,e^{m}_{a}\,\Gamma^{a}\,\Psi-\bar\Psi\left(m_5+{\cal V}(z)\right)\,\Psi\right],
\end{equation}
with $m_5$ a 5D constant fermion bulk mass and ${\cal V}(z)=\kappa^{2}\,z^{2}/R$  is an effective potential chosen to lead to exact solutions for the fermionic spectra. One can think of a dressed mass in this model given by $m_5 + {\cal V}(z)$. The covariant derivative is given by 
\begin{equation}
    \mathcal{D}_{m}\equiv \partial_{m}+\frac{1}{2}\omega_{m}^{bc}\,\Sigma_{bc},
\end{equation}
where $\Sigma_{bc}=\frac{1}{4}[\Gamma_{b},\Gamma_{c}]$, with Dirac matrices $\Gamma^{a}=(\gamma^{\mu},-i\gamma^{5})$. The factor $e^{a}_{n}$ is the vielbein which couples the fermion to the  $AdS_{5}$ space, 
\begin{equation}
    e^{m}_{a}=\frac {z} {R} \,\delta^{m}_{a},
\end{equation}
with $m=0, 1, 2, 3, 5$, and $\omega_{m}^{bc}$ is the spin connection, 
\begin{equation}
\omega^{a\,b}_{m}=- \frac {R}{z} 
\left( \delta^a_z \delta^b_m - \delta^b_z \delta^a_m
\right). 
\end{equation}
After a rescaling of the fermionic field
\begin{equation}\label{Psi}
    \Psi(x^{\mu}, z)
    =  
    e^{\kappa^{2}\,z^{2}/2}\psi(x^{\mu}, z),
\end{equation}
the action reads
\begin{equation}
    S=\int dz\,d^{4}x\,\sqrt{-g}\,\left[\frac{i}{2}\bar\psi\,e^{m}_{a}\,\Gamma^{a}\,\mathcal{D}_{m}\psi-\frac{i}{2}(\mathcal{D}_{m}\,\psi)^{\dagger}\,\Gamma^{0}\,e^{m}_{a}\,\Gamma^{a}\,\psi-\bar\psi\left(m_5+\frac{\kappa^{2}z^{2}}{R}\right)\,\psi\right].
\end{equation}
Now, considering a field decomposition into left and right chiral components
\begin{equation}
    \psi(x^{\mu}, z)=\psi_{_L}(x^{\mu}, z)+\psi_{_R}(x^{\mu}, z),
\end{equation}
one can write each chiral mode in terms of the four-dimensional Dirac spinors $ u_s (P)$ as
\begin{equation}
    \psi_{\frac{L}{R}}(x^{\mu}, z)= e^{i P\cdot x} \frac 1 2 (1 \mp \gamma^5) u_s (P) \frac{z^2}{R^2}{f}_{\frac{L}{R}}(z), 
\end{equation}
where $P^\mu$ is the four-dimensional momenta of the hadron. The equations of motion can be written as 
\begin{equation}
    \left[-\partial_{z}^{2}+\kappa^{4}\,z^{2}+2\,\kappa^{2}\left(R m_5\mp\frac{1}{2} \right)+\frac{R m_5(R m_5\pm 1)}{z^{2}} \right]{f}^{n}_{\frac{L}{R}}(z)=M_{n}^{2}\,{f}^{n}_{\frac{L}{R}}(z),
\end{equation}
where $M_n$ are the hadron masses in four dimensions.  
In the interesting case where $R m_5\ge 1/2$, the corresponding spectrum reads (for the left and right modes, equally): 
\begin{equation}
    M_{n}^{2}=4\,\kappa^{2}\left(n+R m_5+\frac{1}{2}\right);\qquad \qquad  (n=0, 1, 2, \dots).\label{MSWM}
\end{equation}
 The normalized solutions to the equations of motion are the analytical functions given by
\begin{align}
    {f}^{n}_{_{L/R}}(z)&=\sqrt{\frac{2\,\Gamma(n+1)}{\Gamma(n+p_\pm+1)}}\, |\kappa|^{p_\pm+1}
    \,z^{p_\pm+1/2}\,
    e^{-\kappa^{2}z^{2}/2}\,
    L^{p_\pm}_{n}(\kappa^{2}z^{2}),  
\end{align}
where $L^{p_\pm}_{n}(\kappa^2\,z^{2})$ are the associated Laguerre polynomials with $p_{\pm}=|R\, m_5\pm 1/2|$ for the left and right modes, respectively.}

{So, we see that the soft-wall and the deformed AdS models are different since the first gives rise to analytical solutions for fermions while the latter implies numerical solutions, as we saw in \ref{HM} (see also \cite{Nascimento:2023dzx}). An analogous situation happens for the case of the graviton with analytical solutions in the soft-wall \cite{Abidin:2009hr,Mamo:2019mka} and our numerical solutions in the deformed AdS model presented in \ref{GFFs}.}

%\bibliographystyle{unsrt}
% With title, without hyperlinks, in order of appearance in text. 

\bibliography{references.bib}

\end{document}